\documentclass{emulateapj}

\newcommand{\gsim}{\raisebox{0.3mm}{\em $\, >$} \hspace{-3.3mm}
\raisebox{-1.8mm}{\em $\sim \,$}}
\newcommand{\bm}[1]{\mbox{\boldmath $#1$}}

\slugcomment{To Appear in the Astrophysical Journal}
\shorttitle{SUPERCRITICAL DISK ACCRETION}
\shortauthors{OHSUGA \& MINESHIGE}

\begin{document}


\title{Why Is Supercritical Disk Accretion Feasible?}


\author{Ken Ohsuga}
\affil{Department of Physics, Rikkyo University, 
Toshimaku, Tokyo 171-8501, Japan; \\
and Institute of Physical and Chemical Research (RIKEN), 
2-1 Hirosawa, Wako, Saitama, 351-0198, Japan}

\and
\author{Shin Mineshige}
\affil{Yukawa Institute for Theoretical Physics,
Kyoto University, Kyoto 606-8502, Japan}

\begin{abstract}
Although the occurrence of steady supercritical disk accretion 
onto a black hole has been speculated about since the 1970s, 
it has not been accurately verified so far. 
For the first time, we previously demonstrated it 
through two-dimensional, long-term 
radiation-hydrodynamic simulations. To clarify why this accretion is 
possible, we quantitatively investigate the dynamics 
of a simulated supercritical accretion flow with
a mass accretion rate of $\sim 10^2 L_{\rm E}/c^2$
(with $L_{\rm E}$ and $c$ being, respectively, the Eddington 
luminosity and the speed of light). We confirm two important 
mechanisms underlying supercritical disk accretion flow,
as previously claimed,
one of which is the radiation anisotropy arising from 
the anisotropic density distribution 
of very optically thick material. 
We qualitatively show that
despite a very 
large radiation energy density, 
$E_0\gsim 10^2L_{\rm E}/4\pi r^2 c$
(with $r$ being the distance from the black hole),
the radiative flux $F_0\sim cE_0/\tau$ could be small 
due to a large optical depth, typically $\tau\sim 10^3$, in the disk.
Another mechanism is photon trapping, quantified by 
${\bm v}E_0$, where ${\bm v}$ 
is the flow velocity.  
With a large $|{\bm v}|$ and $E_0$, this term 
significantly reduces the radiative flux and even makes it 
negative (inward) at $r < 70 r_S$,  where $r_S$ is
the Schwarzschild radius.
Due to the combination of these effects, 
the radiative force in the direction along the disk plane is largely 
attenuated so that the gravitational force barely exceeds the sum 
of the radiative force and the centrifugal force. As a result, 
matter can slowly fall onto the central black hole mainly along the 
disk plane with velocity much less than the free-fall velocity,
even though the disk luminosity exceeds the Eddington 
luminosity. Along the disk rotation axis, in contrast,
the strong radiative force drives strong gas outflows.
\end{abstract}


\section{Introduction}

Recently, very bright objects which may be undergoing
supercritical (or super-Eddington) accretion flows
have successively been found.
Good examples are ultraluminous X-ray sources 
(ULXs; \citeauthor{WMM01} \citeyear{WMM01};
\citeauthor{Ebisawa03} \citeyear{Ebisawa03};
\citeauthor{Okajima06} \citeyear{Okajima06}).
These are
pointlike off-center X-ray sources whose X-ray luminosity significantly 
exceeds the Eddington luminosity of a neutron star \citep{Fabbiano89}.
Because of substantial variations, it is reasonable to assume
that the ULXs are 
single compact objects powered by accretion flows \citep{Makishima00}.
If so, there are two possibilities to account for 
large luminosities exceeding 
the Eddington luminosity for a mass of 100 $M_\odot$:
sub-critical accretion onto an intermediate-mass black hole (IMBH) 
and supercritical accretion onto a stellar-mass black hole.
We support the latter possibility, 
since through the fitting to several {\it XMM-Newton} EPIC data of ULXs, 
which have been claimed as good IMBH candidates,
we have found evidence of supercritical flows 
\citep{Vierdayanti06}.
Another interesting group is narrow-line Seyfert 1 galaxies 
\citep[see][for a review]{Boller04}.
Because of their relatively small black hole masses,
they have in general large Eddington ratios 
($L/L_{\rm E}$ with $L$ and $L_{\rm E}$ being the 
luminosity and the Eddington luminosity, respectively),
and some of them seem to fall in the slim-disk regimes
(\citeauthor{Mineshige00} \citeyear{Mineshige00};
\citeauthor{Kawaguchi03} \citeyear{Kawaguchi03};
see also \citeauthor{Wang99} \citeyear{Wang99})

Despite growing evidence indicating the existence of supercritical
accretion flows in the universe, 
theoretical understanding is far from being complete.
It is well known that any spherically accreting object, 
irrespective of the nature of the central source, 
cannot emit above the Eddington luminosity,
since otherwise significant radiative force 
will prevent accretion of the gas. 
If we examine detailed radiation-matter interactions in the interior,
however, 
we notice that the situation is not so simple.   Actually,
radiation produced at the very center cannot immediately reach
the surface (i.e., photosphere), since photons generated at
the center should suffer numerous scatterings with accreting material
and thus take a long time to reach the surface.
If the matter continuously falls, and if the mass accretion timescale
is shorter than the mean travel time for photons to reach the surface
(the diffusion timescale), photons at the core may  not be able to go out.  
This is the so-called photon-trapping effect
\citep[e.g.,][]{Begelman78,HC91}.
Here we define the trapping radius,
inside which radiation-matter interaction is so frequent that
photons are trapped within the accretion flow.
Inside this trapping radius, therefore,
radiative flux can be negative (i.e., inward).
Thus, the apparent luminosity is reduced
as compared with the case without photon trapping.
Nevertheless, the concept of the Eddington luminosity is still valid,
since far outside the trapping radius 
radiative flux should be outward 
and its absolute value should be less than
$L_{\rm E}/4\pi r^2$,
where $r$ is the distance from the central black hole,
in the quasi-steady state.
Radiation-hydrodynamic (RHD) simulations of spherically symmetric
supercritical accretion flows
have been performed by \citet{BK83}.

The situation may differ in the case of disk accretion, 
since the radiation 
field is not isotropic due to inhomogeneous matter distribution.
That is, matter can fall predominantly along the disk plane,
whereas radiation can go out along its rotating axis, where
matter density is low.
In other words, 
the main directions of the inward matter flow and 
the outward radiative flux are not parallel to each other
in disk accretion, leading to a situation in which
the radiative force does not completely
counteract the gravitational force.
There is room for the possibility of a supercritical accretion flow
with super-Eddington luminosity.

Based on such an argument, 
many researchers have speculated about
the occurrence of supercritical flows in disk accretion systems.
\citet{SS73} discussed the possibility of supercritical disk accretion 
based on a one-dimensional steady model
\citep[see also][]{MRT76}. They mentioned that the mass accretion rate
would not be steady but oscillate if the mass accretion rate
exceeded the critical rate; otherwise, a part of the accreting matter 
might be ejected from the disk as a disk wind. 
Radiatively driven outflows from supercritical disks 
were investigated by \citet{Meier79}, \citet{JR79}, and \citet{Icke80}.

Despite a long history in the study of 
supercritical disk accretion flows, the occurrence 
of steady supercritical disk accretion has not yet been
accurately verified.
Similar simulations have been performed since the 1980s, but all of them 
calculated only the initial transient phase.  
Their conclusions then were not general, since the back-reactions
(i.e., enhanced radiation pressure), which may inhibit steady 
flow, were not accurately evaluated.
Although the radiative force predominantly has an effect in the vertical
direction, it should also have an effect on the material within the disk.
Hence, in the direction parallel to the disk plane, 
the situation may be similar to or 
more severe than the case of spherical accretion.
This is because the radiative force, 
together with the centrifugal force,
may possibly overcome the gravitational force.
To study the possibility of supercritical disk accretion flows,
precise and quantitative research 
treating both supercritical disks and outflows, 
and which also takes into consideration 
multi-dimensional effects, is needed.

\citeauthor{O05} (\citeyear[][hereafter Paper I]{O05})
have confirmed the occurrence of
quasi-steady supercritical disk accretion onto a black hole
by two-dimensional RHD simulations.
The motivation of the present study is to
investigate the physical mechanisms
which make supercritical disk accretion possible.
For this purpose we examine quantitatively 
the flow motion and force fields 
via the radiative flux, rotation, and gravity
of a supercritical disk accretion flow
onto a central black hole,
based on the two-dimensional RHD simulation data from Paper I.
Through detailed inspection of the results, 
it will be possible to
clarify the physics behind supercritical disk accretion flows.
In \S 2 we plot the spatial distributions of
several key quantities which control flow dynamics.
A discussion is given in \S 3.

\section{Dynamics of a Supercritical Disk Accretion Flow}

\subsection{Basic Considerations}

Here we adopt the spherical coordinates $(r,\theta,\varphi)$,
with $\theta=\pi/2$ corresponding to the disk plane
and the origin being set at the central black hole.
The momentum equation for the radial component of the flow is
\begin{equation}
 \frac{d v_r}{dt} = 
  f_{\rm grav} + f_{\rm rad}^r + f_{\rm pres}^r
  + f_{\rm cent}^r.
\end{equation}
Here  $v_r$ is the radial component of the flow velocity,
$d/dt \equiv \partial/\partial t + v_r \partial/\partial r$ 
is the Lagrangian derivative,
$f_{\rm grav}$ is the gravitational force by the central black hole,
$f_{\rm rad}^r = \chi F_0^r/c$ 
is the radiative force
with $\chi$ being absorption and scattering coefficients,
$c$ being the speed of light,
and $F_0^r$ representing the radial component of 
the radiative flux in the comoving frame,
$f_{\rm pres}^r$ is 
the radial component of the gas pressure force,
and $f_{\rm cent}^r = v_\varphi^2/r$ is 
the radial component of the centrifugal force.
We adopt a pseudo-Newtonian potential, thus having
$f_{\rm grav}=-GM/(r-r_S)^2$
with $M$ being the mass of the black hole 
and $r_S \equiv 2GM/c^2$
\citep{PW80}.

To proceed, it is important to distinguish two different views: 
the view from an observer comoving with the accreting gas 
(i.e., the comoving frame)
and the view from an observer standing at infinity (the rest frame).
The $i$th component of radiative flux in the comoving frame, $F_0^i$, 
is related to that in the rest frame, $F^i$, 
on the order of $v/c$,
\begin{equation}
  F^i = F_0^i + v_i E_0 + v_j P_0^{ij},
  \label{flux}
\end{equation}
where $E_0$ and $P_0^{ij}$ are the radiation energy density 
and the radiation stress tensor in the comoving frame, respectively.
Here the final term in equation (\ref{flux}) 
is equal to $v_i E_0/3$ in the optically thick diffusion limit.

In \S 2.3 we show that the gas accretes toward the black hole
in the disk region.
In this region, 
even if the radiative flux in the comoving frame is positive, $F_0^r > 0$ 
(i.e., outward flux), 
$F^r$ can be negative, $F^r < 0$ (inward flux), 
because $v_r < 0$.
The term $v_r E_0$ thus represents photon trapping.
In \S 2.5 we explicitly show this effect in the simulation data.

In contrast, the flow dynamics in the direction perpendicular to
the disk plane is distinct.  In particular,
the centrifugal force has no effect in this direction.
In addition, 
the matter density is significantly smaller along the rotation axis.
Hence, the radiative flux can be much more effective
in the vertical direction, thus driving strong outflows,
as we see in \S 2.3.

\subsection{Overview of the Simulated Flow}

\begin{figure*}
 \epsscale{1.075}
 \plotone{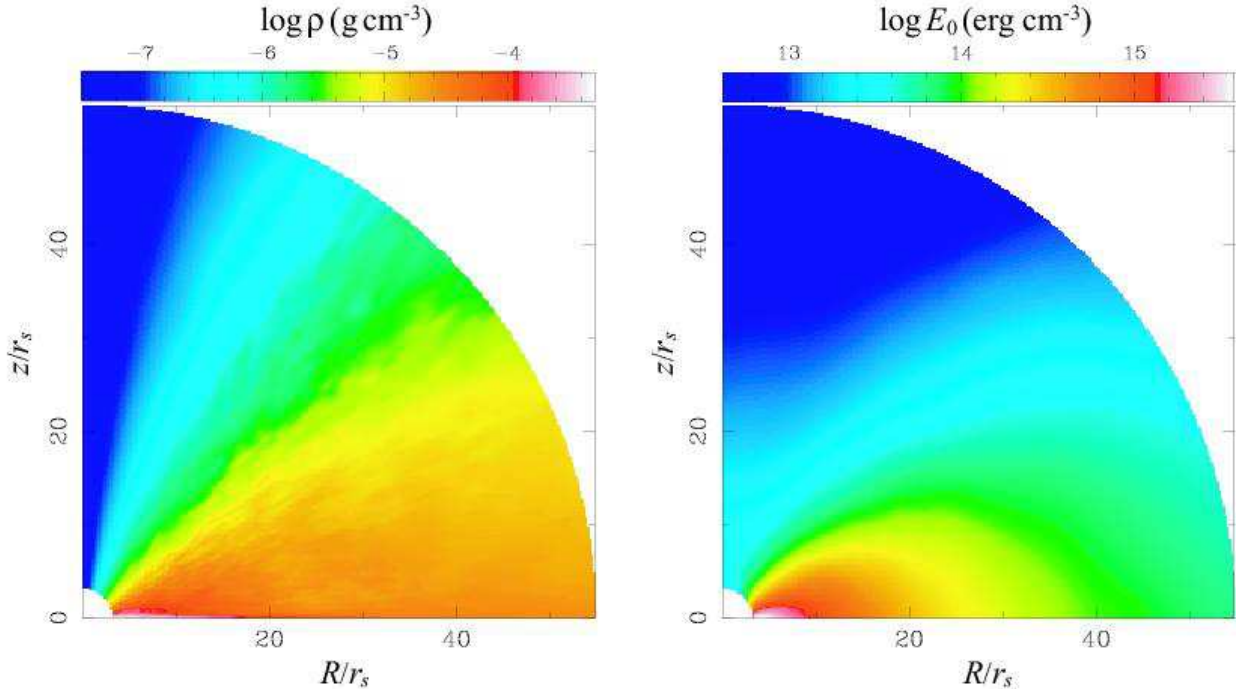}
 \caption{Color contours of matter density ({\it left}) and
 radiation energy density ({\it right}) distributions
 as a function of $R$ and $z$.  Note that
 these values are time-averaged over $t=20-50$ s.
 \label{rhoE}   
 }
\end{figure*}

Our present analysis is 
based on the two-dimensional RHD simulation data from Paper I.
These are the first simulations of supercritical accretion flow
in the quasi-steady state.
Although the research history of such simulations 
stems back to the late 1980s,
when \citet{ECK87} performed numerical simulations for the first time,
their calculations were restricted to the first few seconds 
(see also \citeauthor{Kley89} \citeyear{Kley89};
\citeauthor{OFS97} \citeyear{OFS97};
\citeauthor{KL99} \citeyear{KL99};
\citeauthor{Okuda02} \citeyear{Okuda02}).
Back-reactions were not fully taken into account in their simulations.
In our simulations presented in Paper I,
matter is added continuously from the outer 
disk boundary at $r = 500~r_S$ 
to the initially (nearly) empty space 
at the rate of $10^3 L_E/c^2$.
The injected mass has a small specific angular momentum so that
it first free falls but then forms a rotating disk at 
$r \sim 100 r_S$.
Mirror symmetry is assumed with respect to the disk (equatorial) plane.
The $\alpha$-prescription of the viscosity is employed,
in which the viscosity is set to be proportional to the total pressure
in the optically thick limit and 
the gas pressure only in the optically thin region.
Hence, the viscosity is more effective in the disk region than
in the outflow regions above and below the disk.
The viscosity parameter is assumed to be $\alpha = 0.1$.
Except for the radial-azimuthal component,
all components of the viscous stress tensor are set to be zero.

The radiative transfer was solved under the flux-limited diffusion
approximation \citep{LP81}.
The central object is taken to be a non rotating stellar-mass 
black hole ($M=10 M_\odot$), generating a pseudo-Newtonian potential.
Only $10\%$ of the
inflowing material finally reaches the inner boundary ($3r_S$); 
i.e., $90\%$ of the mass input gets stuck in the dense, 
disk like structure around the equatorial plane,
or transforms into the known collimated high-velocity outflows 
perpendicular to the equatorial plane
or into low-velocity outflows with wider opening angles.
The resulting luminosity is about 
3 times larger than the Eddington luminosity.

Figure \ref{rhoE} indicates the time-averaged contours of the matter density
({\it left panel}) and the radiation energy density ({\it right panel})
on the $R$-$z$ plane ($R=r \sin \theta$ and $z=r \cos \theta$),
where both axes are normalized 
by the Schwarzschild radius ($r_S$).
Note that these panels are similar to Figures 3 and 5 of Paper I,
except that Figure \ref{rhoE} of the present paper has contours which
are time-averaged over $t=20-50$ s,
whereas those in Paper I are snapshots at $t=10$ s 
(i.e., in a quasi-steady phase).
The high concentration of gas and radiation in the region near the 
black hole on the equatorial plane is clear.


\subsection{Dynamics of the Simulated Flow}
\begin{figure}[b]
 \epsscale{1.075}
 \plotone{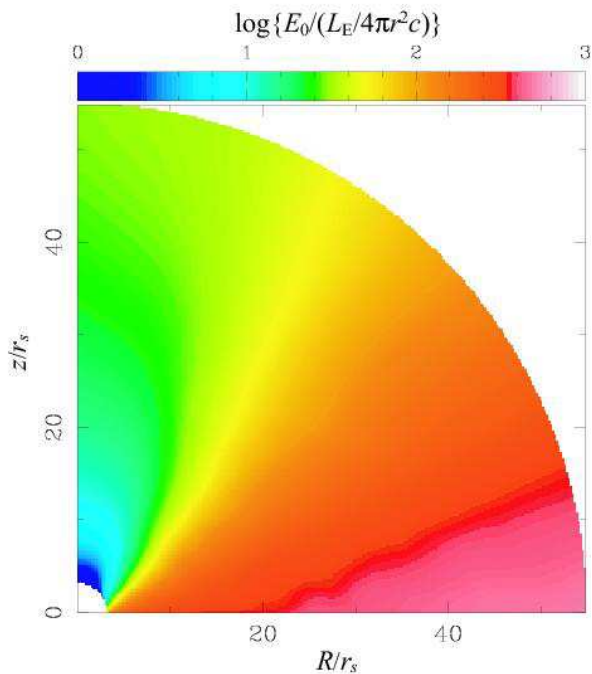}
 \caption{Same as Fig. 1, but with the radiation energy density
 normalized by $L_{\rm E}/4\pi r^2 c$.
 \label{E0}   
 }
\end{figure}

First we plot the time-averaged 
radiation energy density normalized by
$L_{\rm E}/4\pi r^2 c$ in Figure \ref{E0}.
\begin{figure*}
 \epsscale{1.075}
 \plotone{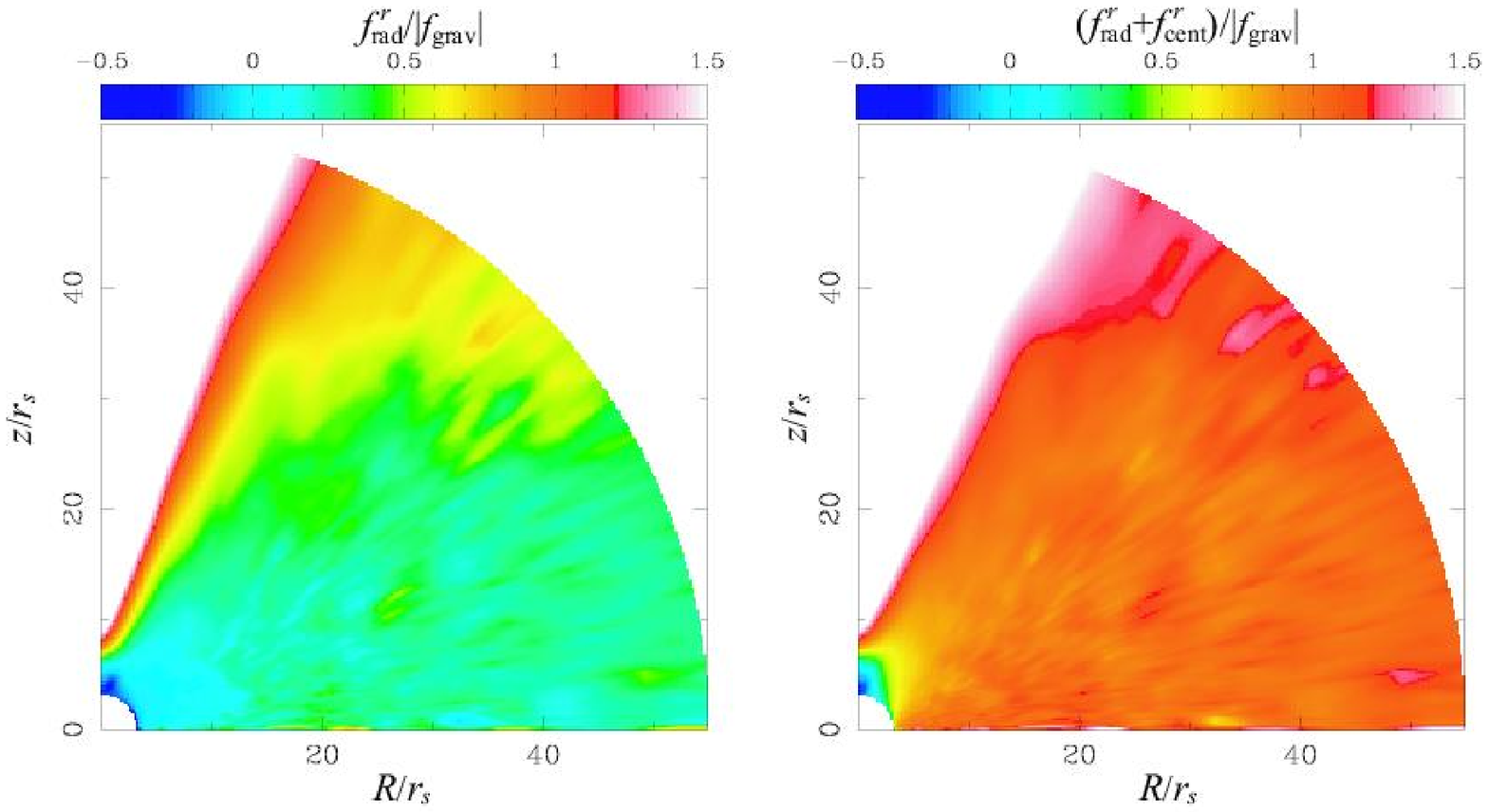}
 \caption{Same as Fig. 1, but with the color contours
 of the ratio of the radial component of the radiative force 
 to the gravitational force ({\it left}) 
 and that of the sum of the radiative
 force and the centrifugal force to the gravitational force ({\it right}).
 \label{2DForce}   
 }
\end{figure*}
We find that this value is larger than unity in most regions.
In particular, we have 
$E_0 \gsim 10^2L_{\rm E}/4\pi r^2 c$ in the disk region.
This means that there exists so large a number of photons 
that if we simply assume $F_0\sim cE_0$
as in the optically thin region,
the radiative force should highly exceed the gravity,
which inevitably prevents the inflow motion.
The simulation data, however, do show inflow motion (discussed below).

Figure \ref{2DForce} illustrates the time-averaged 
force balance in the radial ($r$)
directions with various $\theta$-values from the origin.
(It should be noted that since force is a vector, we need to specify
the direction to draw such contours.  Here
we only plot the radial component of the forces.)
Figure \ref{2DForce} ({\it left})
represents the ratio of the radial component 
of the radiative force
to the gravitational force.  
Especially surprising is that
most regions are turquoise, indicating that
the radiative force is outward and less than the gravitational force.
Why is the radiative force so weak, 
even though the radiation energy density is very large, 
$E_0 \gsim 10^2L_{\rm E}/4\pi r^2c$ 
(see Fig. \ref{E0}),
and even though the total luminosity exceeds the Eddington luminosity, 
$L > L_{\rm E}$?
This is because the disk is very optically thick.
Although the absolute value of the radiative flux is equal
to $cE_0$ in the optically thin limit, 
it is roughly given by $cE_0/\tau$, and hence, $F_0$ can be
very small, $F_0\ll cE_0$,
in the optically thick region ($\tau\gg 1$),
where $\tau$ is the optical thickness.
Approximately, we find 
$\tau=\rho \chi r \sim 10^3
(\rho/10^{-4.5}{\rm g\,cm^{-3}})(r/30r_S)$,
where $\rho$ is the gas density.
Thus, flux is attenuated a great deal.  
We thus have a relatively small radiative flux within the disk
in spite of a large $E_0$
(see \S 2.4 for more detail).
There are two exceptions:
the color of the region
in the vicinity of the black hole is blue, which means
that the radiative force is inward.  
In this region, a large number of photons 
are swallowed by the black hole, $F_0^r<0$,
so that the gas is accelerated inward.
The region near the rotation axis, in contrast, has 
a white color, which means
that the radiative force exceeds the gravitational force.

To discuss how matter flows, we also need to consider the centrifugal
force.  (Note that the gas pressure is negligible in the present case.)
Figure \ref{2DForce} ({\it right}) illustrates the ratio of the sum of 
the radial components of the radiation and centrifugal forces 
to the gravitational force.
Certainly, the repulsing force is enhanced
along the disk plane. However, this force ratio is still around unity
in most regions
except in the very vicinity of the black hole.
That is, the gas is nearly in balance of forces in wide regions.
The regions near the rotation axis again have white colors, which means
that the radiative force dominates, as in Figure \ref{2DForce} ({\it left}) 
(Note that the centrifugal force is not effective near the rotation axis.)

\begin{figure}
 \epsscale{1.075}
 \plotone{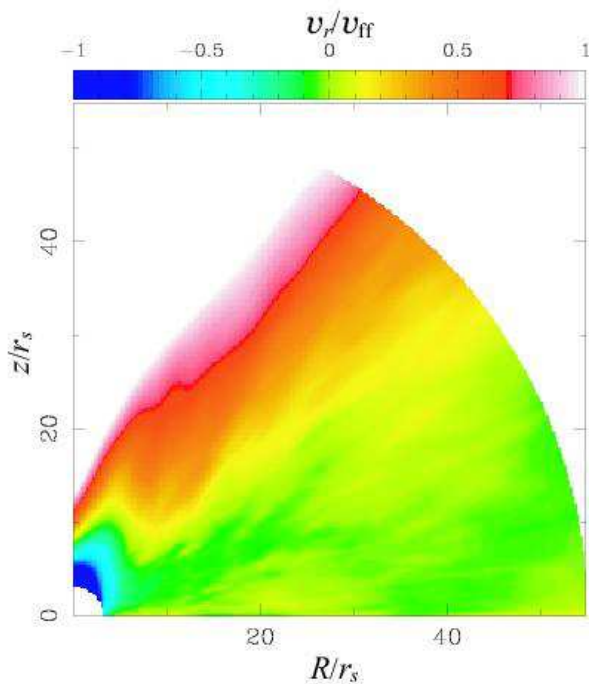}
 \caption{Same as Fig. 1, but with the radial velocity 
 normalized by the free-fall velocity.
 \label{vr}   
 }
\end{figure}

Since there are regions with a negative effective force 
in the radial direction,
we expect regions with negative velocity (i.e., inflow).
Figure \ref{vr} shows the spatial distribution of
the radial component of the matter velocity
in units of the free-fall velocity.
Interestingly, the flow is inward in the disk region,
but the infall velocity is much smaller than the
free-fall velocity.
Along the rotation axis, conversely,
the flow is outward, and its magnitude is 
larger than the free-fall velocity.

\begin{figure}
 \epsscale{1.15}
 \plotone{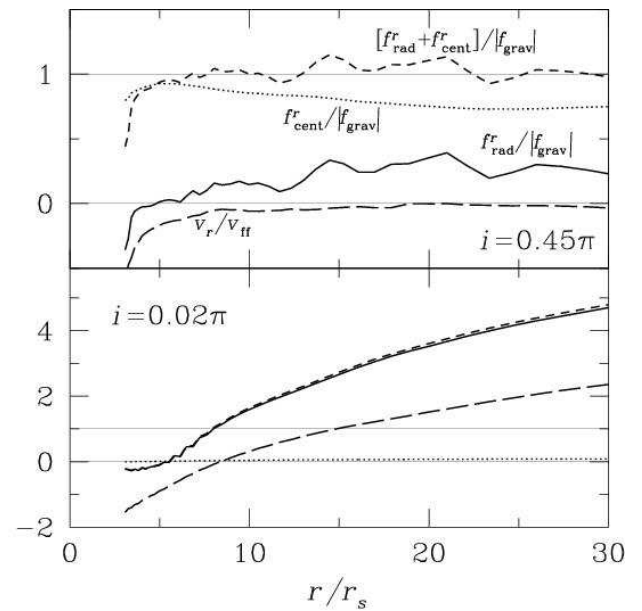}
 \caption{time-averaged one-dimensional profiles of 
 the radiative force ({\it solid lines}),
 the centrifugal force ({\it dotted lines}),
 their sum ({\it dashed lines}), 
 and the radial velocity ({\it long-dashed lines}) 
 for the case with a large
 polar angle, $i=0.45 \pi$ ({\it top}), and
 with a small angle, 
 $i=2.3\times 10^{-2} \pi$ ({\it bottom}).
 Note that all the forces
 are normalized by the gravitational force,
 while the radial velocity is normalized by the free-fall velocity.
 In both panels the two thin horizontal lines indicate
 the values of 0 and 1.
 \label{1DForce}   
 }
\end{figure}

Let us have a more quantitative discussion based on the 
one-dimensional distributions of these quantities, since
it is difficult to obtain their precise values from the color contours.  
We calculate several quantities along two radial lines:
one at 0.45$\pi$ (close to the equatorial plane)
and the other at $\theta = 2.3 \times 10^{-2} \pi$ 
(nearly along the rotation axis).
They are illustrated in the top and bottom panels of Figure
\ref{1DForce}, respectively.

We see in the top panel that the radiative force is not
negligible, although it is smaller than the centrifugal force.
The sum of the radiative and centrifugal forces
is nearly balanced with the gravitational force.
Radiation does work as a ``radiation cushion,'' which
decelerates the accretion of the gas 
in cooperation with the rotation (centrifugal force).
Deceleration via radiative flux was 
also reported by \citet{BK80, BK83}.
Figure \ref{1DForce} ({\it top}) also shows that
the radial velocity is much smaller than the free-fall velocity
and only barely below zero (inflow),
although there also exist regions with $v_r\sim 0$ (no inflow).

In contrast, Figure \ref{1DForce} ({\it bottom}) shows that the ratio of
the radiative force to the gravitational force grows as
the distance increases.  The outward velocity thus grows rapidly upward.
We confirm again that 
the strong outflow around the rotation axis 
is produced by the strong radiative force.
However, the velocity is negative in the very vicinity of the hole.
This means that the radiation which produces the huge radiative
force in the vertical direction does not originate from the region 
just outside the event horizon but from the region 
with $E_0$ at maximum [at $(5-10)r_S$].

Here we note that the angular momentum is very small around 
the rotation axis, leading to a negligible centrifugal force 
({\it dotted line}). This is because a
significant amount of material exists 
even inside the radius of the marginally stable orbit, $R<3r_S$,
in the supercritical flow,
so the matter can lose angular momentum there
\citep{KFM98, WM03}.
This situation contrasts with that of the subcritical flow, in which
there is a sharp density drop at $R=3 r_S$.

Our computational domain covers exactly down to the 
rotation axis. Numerical simulations show that most of the outflow 
matter is blown away into an oblique direction by the radiative
force in cooperation with the centrifugal force.


\subsection{Geometric Effect}
Here we consider the reason why the radiative force 
in the radial direction
does not dominate over the gravitational force inside the disk,
even though $E_{\rm 0}$ is larger than $10^2L_{\rm E}/4\pi r^2 c$
as shown in Figure \ref{E0}.
The radial component of the radiative flux is evaluated as 
$F_0^r \sim \left(c/3\rho \chi \right) 
\partial E_0/\partial r$
in the disk region. 
It is attenuated since the disk is dense and 
very optically thick ($\tau\gg 1$).
That is, even though the radiation energy density itself 
is large, the high matter density significantly suppresses 
the radiative flux inside the disk.
%
%
However, the high density (large optical thickness) 
is not the only condition for the occurrence of supercritical accretion.
In the case of a spherical system, 
the radiative flux should be $L/4\pi r^2$
even if the matter is very optically thick.
The radiative force dominates 
over the gravitational force when $L>L_E$,
preventing the accretion of matter.
Hence, there should be another reason that
supercritical disk accretion is realized
in spite of $L>L_E$.

Whereas the radiation field is isotropic in the spherical case by definition,
it can be anisotropic in the disk case.
Indeed, the simulated profile of the radiation energy density 
is highly anisotropic (Fig. \ref{rhoE}, {\it right}).
Unlike in spherical geometry,
photons can escape from the less dense region
around the rotation axis 
without thrusting through the dense disk region.
Thus, the effective radiative force is attenuated
in the disk region.
Here we note that,
even in the presence of an anisotropic matter distribution, $L$ cannot
exceed $L_{\rm E}$ if the medium is optically thin.
To sum up, supercritical disk accretion with super-Eddington luminosity
is realized due to a large optical thickness and anisotropy 
of the radiation field.

\subsection{Photon-trapping Effects}
\begin{figure}[b]
 \epsscale{1.2}
 \plotone{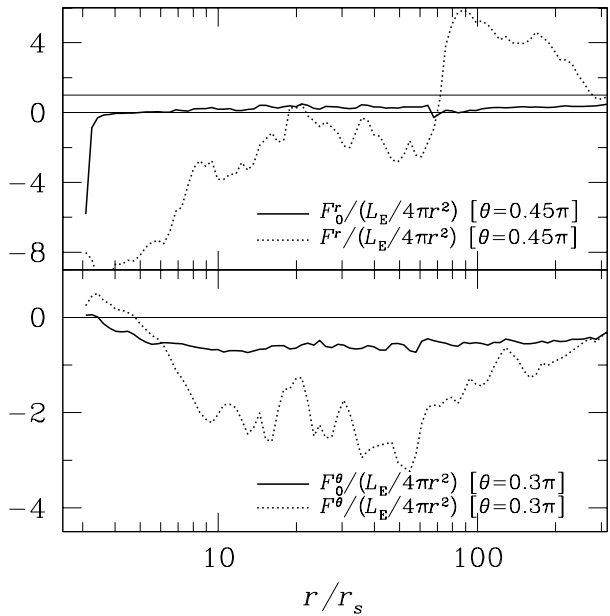}
 \caption{
 {\it Top}: Radial components of the 
 time-averaged radiative fluxes in the comoving
 frame ($F_0^r$; {\it solid line}) and the rest frame 
 ($F^r$; {\it dotted line})
 near the equatorial plane as a function of the radius.
 {\it Bottom}: Same as the top panel, but with the polar components
 of the radiative fluxes in the comoving frame, $F_0^{\theta}$ 
 ({\it solid line}),
 and in the rest frame, $F^{\theta}$ ({\it dotted line}) along a line with
 $\theta = 0.3 \pi$. 
 \label{flux2}   
 }
\end{figure}

Photon-trapping effects assist the occurrence of supercritical
accretion.
Figure \ref{flux2} ({\it top}) indicates the radiative fluxes in
the comoving and rest frames inside the disk.
As shown in this figure,
the radial component of
the radiative flux in the rest frame ({\it dotted line}) is negative
in the region of $r<70r_S$,
whereas that in the comoving frame ({\it solid line}) is positive
except in the vicinity of the black hole.
This means that the radiation energy is transported inward
via photon trapping, $v_r E_0<0$ (see equation [\ref{flux}]).
Some part of this radiation energy is swallowed by the black hole,
whereby the luminosity is reduced.
This effect attenuates the radiative force,
supporting the accretion.

The anisotropy of the radiation field is enhanced 
by photon trapping.
This is another important role of the photon trapping.
The radiation energy, which is advected inward with the matter
by the photon trapping
but not swallowed by the black hole, 
is transported to the vertical 
direction and finally released from the disk surface.
It is shown in Figure 5 ({\it bottom}), which plots 
the $\theta$-components of the radiative fluxes in the comoving 
({\it solid line}) and the rest ({\it dotted line}) frames 
along a radial line at
$\theta=0.3\pi$, that the entire region is below the photosphere.
In this figure we can see that the radiation energy 
is transported toward the disk surface, $F^{\theta}<0$,
in most of the region.
We also find $F^{\theta}<F_0^{\theta}$ in this figure.
This implies that the gas motion
contributes to the transport of radiation energy 
toward the disk surface, $v_\theta E_0<0$.
To conclude, the photon-trapping effect
enhances the anisotropy of the radiation field,
assisting the occurrence of supercritical accretion.



Here we note that the gas motion in the vertical direction
seems to be driven by convection in cooperation with 
the radiative force.
In fact, we find that the adiabatic conditions
for radiation-pressure-dominated matter
are roughly satisfied along the vertical line in the disk region,
$\partial\ln P/\partial\ln \rho \sim 4/3$
and $\partial\ln T/\partial\ln P \sim 1/4$,
where $P$ is the total pressure and $T$ is the temperature.

\section{Discussion}
\subsection{Structure of the Supercritical Disk Accretion Flow}
\begin{figure*}
 \epsscale{0.87}
 \plotone{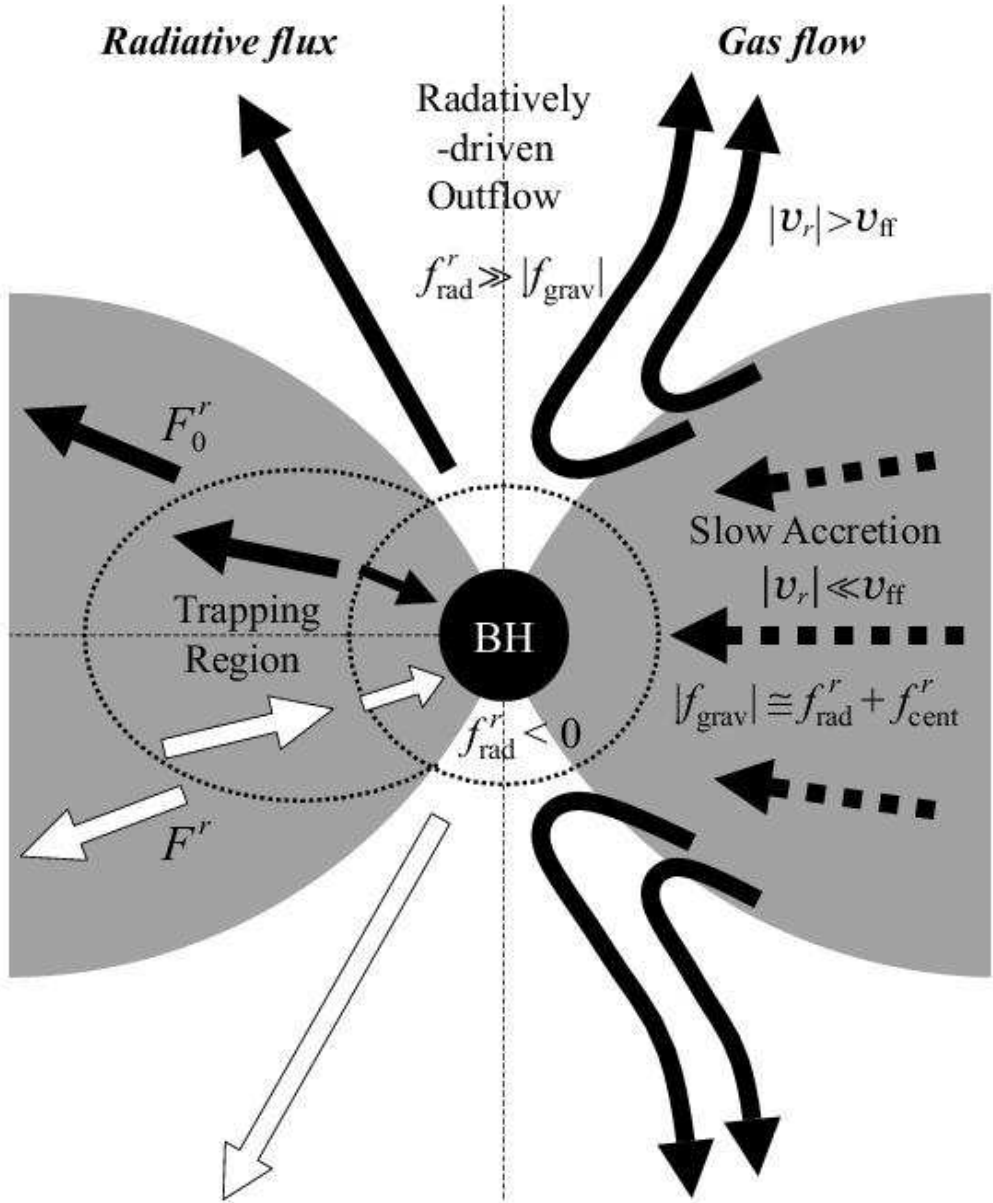}
 \caption{Schematic picture of the supercritical disk
 accretion flow around a black hole.
 The gas motion is shown on the right: the high-velocity outflow
 ({\it solid arrows}) and the slow accretion flow 
 ({\it dashed arrows}). The left side indicates 
 the radiative fluxes in the comoving frame ({\it black arrows}) and 
 the rest frame ({\it white arrows}).
 \label{pic}
 }
\end{figure*}

We summarize our simulation results in a schematic
picture (see Fig. \ref{pic}). 
In this figure
the radiative flux in the comoving frame, $F_0^r$, is
positive (outward) except in the vicinity of the black hole.
However, the radiative flux in the rest frame, 
$F^r (\sim F_0^r+4v_r E_0/3)$, 
is negative (inward) via photon trapping, $v_r E_0<0$,
in the trapping region.
In the vicinity of the black hole,
both $F_0^r$ and $F^r$ are negative.
Matter slowly accretes inside the disk,
since the sum of the radiative and 
centrifugal forces is nearly balanced with the 
gravitational force.
Here we stress again that 
the radiative force counteracts the gravitational force
in spite of the trapping region because $F_0^r>0$.
Radiatively driven high-velocity outflows appear 
above and below the disk.
In the very vicinity of the black hole,
the gas is accelerated inward by the radiative force 
and the gravitational force.

As far as the physical quantities around the equatorial plane
are concerned, the simulated profiles of the density, temperature,
and radial as well as rotational velocities
roughly agree with the prediction of the slim-disk model 
\citep{Abramowicz88}.
Such features have already been shown
in Figure 11 in Paper I.
However, only about $10\%$ of the injected mass can accrete 
onto the black hole, and an almost equal amount of matter is ejected 
as high-velocity outflows. The mass accretion rate
is not constant in the radial direction and decreases near the 
black hole (see Fig. 6 in Paper I).
Thus, we conclude that 
the simulated flows do not perfectly agree with the slim-disk model
with regard to the whole structure of the flow.

\citet{HD07} have investigated local maximum values 
of the accretion rate in the supercritical disk accretion flows.
In their paper
they focused only on the force balance 
in the vertical and radial directions 
around the equatorial plane.
Multi-dimensional effects were not 
taken into consideration.
They revealed that 
the vertical radiative force limits the maximum 
accretion rate at the inner disk region, 
leading to a decrease of the accretion rate
with a decrease of the radius.
Their results imply that 
the disk loses mass 
via the radiatively driven outflows
and the mass accretion rate decreases 
at the inner region.
Such tendencies agree with our results.
As shown in Figure \ref{vr}, our simulations show that 
radiatively driven outflows form above and below the disk.
The mass accretion rate decreases with 
a decrease of the radius
(see Fig. 6 in Paper I).

\subsection{Dependency of the Mass Accretion Rate}
In the present study, focusing on numerical simulations
in which the mass input rate at the outer boundary
is set to be $\dot{M}_{\rm input}=10^3L_{\rm E}/c^2$,
we show that 
the radiative force is attenuated in the disk region
via large optical thickness,
which makes supercritical disk accretion possible.
Such dilution of the radiative force would effectively operate
even if the mass input rate (mass accretion rate) varied.
In fact, 
simulations with mass input rates 
of $3\times 10^2L_{\rm E}/c^2$
and $3\times 10^3L_{\rm E}/c^2$
show that 
$E_0/\rho$ is almost independent 
of the mass input rate,
although the radiation energy density goes up 
as the mass input rate increases.
That is, the dynamics is not sensitive to the precise value of 
$\dot{M}_{\rm input}$,
as long as it largely exceeds the critical value.

So far, we have studied steady accretion flows.
Although a highly supercritical disk 
($\dot{M}_{\rm input}>3\times 10^2L_{\rm E}/c^2$)
is quasi-steady,
it has been revealed that 
a moderately supercritical disk
[$\dot{M}_{\rm input}=(10-10^2)L_{\rm E}/c^2$]
is unstable 
and exhibits limit-cycle behavior 
(\citeauthor{O06} \citeyear{O06}, \citeyear{O07};
see also \citeauthor{SH75} \citeyear{SH75};
\citeauthor{SS76} \citeyear{SS76}).
The luminosity goes up and down around the Eddington luminosity.

\citet{O07} has reported that 
the time-averaged mass, momentum, 
and kinetic energy output rates
via the outflow, the mass accretion rate, and the disk luminosity 
increase as the mass input rate increases,
$\propto \dot{M}_{\rm input}^{0.7}
-\dot{M}_{\rm input}^{1.0}$
for $\alpha=0.5$ and 
$\propto \dot{M}_{\rm input}^{0.4}
-\dot{M}_{\rm input}^{0.6}$
for $\alpha=0.1$.

\subsection{Future Work}
As we have already mentioned in \S 3.1,
the sum of the accreting matter and 
the matter ejected as high-velocity outflows
is $20\%$ of the injected mass, and
$80\%$ of the injected matter is 
ejected from the computational domain as low-velocity outflows,
whose velocities do not exceed the escape velocity at the outer 
boundary.
Since such outflowing matter tends to be accelerated by the 
radiative force even at the outside of the computational domain,
it would be blown away from the system.
However, a part of the outflowing matter might return 
to the vicinity of the black hole through the disk region,
since the radiative force does not exceed the gravity
near the equatorial plane.
Numerical simulations with larger computational domains
would make this point clear.

Whereas the resulting mass accretion rate onto the black hole
is around $10^2 L_{\rm E}/c^2$
in the present simulations,
\citet{HD07} have indicated that the mass accretion rate
can increase up to $10^4 L_{\rm E}/c^2$.
They have reported that 
the vertical force balance breaks down via a strong radiative force
if the mass accretion rate exceeds this limit.
However, even in such a case, 
the matter might accrete onto the black hole,
although the strong radiative force would produce
powerful outflows.
To investigate the maximum value of the accretion rate 
is an outstanding issue.
We should perform numerical simulations with larger computational 
domains, since the trapping region is expected to expand 
with the increase of the mass accretion rate.

We reveal in the present paper that photons generated 
deep inside the disk are effectively trapped in the flow, 
leading to supercritical disk accretion.
Although the magnetic fields are not solved in our simulations,
magnetic buoyancy might play an important role in the 
transportation of matter, as well as photons, toward the 
disk surface 
(\citeauthor{Parker75} \citeyear{Parker75};
\citeauthor{SR84} \citeyear{SR84};
\citeauthor{SC89} \citeyear{SC89}).
Magnetic buoyancy might lead to photon generation
near the disk surface if the magnetic fields rise quickly
without dissipation deep inside the disk.
Thus, magnetic buoyancy would dilute the photon-trapping effect. 
A photon bubble instability, which is induced in the magnetized,
radiation-pressure-dominated region, might also suppress
the photon trapping \citep{Begelman02,Turner05}.
In these cases the enhanced radiative force would more effectively 
accelerate the matter around the disk surface,
working to decrease the mass accretion rate.
However, the magnetic fields might prevent such acceleration
if they strongly tie the matter near the disk surface with the disk matter.
In the disk region
the matter might easily accrete toward the black hole,
since the radiation energy density decreases.
Global radiation-magnetohydrodynamic (RMHD) simulations 
would make these points clear.
Local RMHD simulations of accretion flows have been 
performed by \citet{Turner03} and \citet{HKS06}.
In addition, it is thought that disk viscosity has 
magnetic origins 
(\citeauthor{HBS01} \citeyear{HBS01};
\citeauthor{MMM01} \citeyear{MMM01};
for a review see \citeauthor{Balbus03} \citeyear{Balbus03}).
Hence, we should stress again that RMHD simulations are 
very important to more realistically investigate viscous 
accretion flows, although an $\alpha$-viscosity model is 
employed in the present study.

\acknowledgments
We would like to thank the anonymous reviewer for many helpful suggestions.
The calculations were carried out 
by a parallel computer at Rikkyo University
and the Institute of Natural Science, Senshu University.
This work was supported in part 
by a special postdoctoral researchers program in RIKEN (K. O.),
by a research grant from the Japan Society 
for the Promotion of Science (17740111; K. O.),
by Grants-in-Aid from the Ministry
of Education, Science, Culture, and Sports  
(14079205, 16340057; S. M.), 
and by the Grant-in-Aid for the 21st Century COE 
``Center for Diversity and Universality in Physics'' 
from the Ministry of Education, Culture, Sports, 
Science, and Technology of Japan.



\begin{thebibliography}{}
\bibitem[Abramowicz et al.(1988)]{Abramowicz88} 
 Abramowicz, M. A., Czerny, B., Lasota, J. P., \& Szuszkiewicz, E. 1988, 
 \apj, 332, 646
\bibitem[Balbus(2003)]{Balbus03} 
 Balbus, S.~A.\ 2003, \araa, 41, 555 
\bibitem[Begelman(1978)]{Begelman78} 
 Begelman, M. C. 1978, \mnras, 184, 53
\bibitem[Begelman(2002)]{Begelman02} 
 Begelman, M.~C.\ 2002, \apjl, 568, L97
\bibitem[Boller(2004)]{Boller04} 
 Boller, T. 2004, PThPS, 155, 217
\bibitem[Burger \& Katz(1980)]{BK80} 
 Burger, H. L. \& Katz, J. I. 1980, \apj, 236, 921
\bibitem[Burger \& Katz(1983)]{BK83} 
 Burger, H. L. \& Katz, J. I. 1983, \apj, 265, 393
\bibitem[Ebisawa et al.(2003)]{Ebisawa03} 
 Ebisawa, K., Zycki, P., Kubota, A., Mizuno, T., \& Watarai, K.
 2003, \apj, 597, 780
\bibitem[Eggum et al.(1987)]{ECK87} 
 Eggum, G. E., Coroniti, F. V., \& Katz, J. I. 1987, \apj, 323, 634
\bibitem[Fabbiano(1989)]{Fabbiano89} 
 Fabbiano, G. 1989, \araa, 27, 87
\bibitem[Hawley et al.(2001)]{HBS01} 
 Hawley, J. F., Balbus, S. A. \& Stone, J. M. 2001, \apj, 554, L49
\bibitem[Heinzeller \& Duschl(2007)]{HD07} 
 Heinzeller, D., \& Duschl, W.~J.\ 2007, \mnras, 374, 1146 
\bibitem[Hirose et al.(2006)]{HKS06} 
 Hirose, S., Krolik, J.~H., \& Stone, J.~M.\ 2006, \apj, 640, 901 
\bibitem[Houck \& Chevalier(1991)]{HC91} 
 Houck, J. C. \& Chevalier, R. A. 1991, \apj, 376, 234
\bibitem[Icke(1980)]{Icke80} 
 Icke, V. 1980, \aj, 85, 329
\bibitem[Jones \& Raine(1979)]{JR79} 
 Jones, B. C. \& Raine, D. J. 1979, \aap, 76, 179
\bibitem[Kato et al.(1998)]{KFM98} 
 Kato, S., Fukue, J., \& Mineshige, S. 1998,
 Black-Hole Accretion Disks (Kyoto: Kyoto Univ. Press)
\bibitem[Kawaguchi(2003)]{Kawaguchi03} 
 Kawaguchi, T. 2003, \apj, 593, 69
\bibitem[Kley(1989)]{Kley89}
 Kley, W. 1989, \aap, 222, 141
\bibitem[Kley \& Lin(1999)]{KL99}
 Kley, W. \& Lin, D. N. C. 1999, \apj, 518, 833
\bibitem[Levermore \& Pomraning(1981)]{LP81} 
 Levermore, C. D. \& Pomraning, G. C. 1981, \apj, 248, L321
\bibitem[Makishima et al.(2000)]{Makishima00} 
 Makishima, K., et al. 2000, \apj, 535, 632
\bibitem[Machida et al.(2001)]{MMM01} 
 Machida, M., Matsumoto, R., \& Mineshige, S. 2001, \pasj, 53, L1
\bibitem[Maraschi et al.(1976)]{MRT76} 
 Maraschi, L., Reina, C., \& Treves, A. 1976, \apj, 206, 295
\bibitem[Meier(1979)]{Meier79} 
 Meier, D. L. 1979, \apj, 233, 664
\bibitem[Mineshige et al.(2000)]{Mineshige00} 
 Mineshige, S., Kawaguchi, T., Takeuchi, M., \& Hayashida, K. 2000,
 \pasj, 52, 499
\bibitem[Ohsuga(2006)]{O06} Ohsuga, K. 2006, \apj, 640, 923
\bibitem[Ohsuga(2007)]{O07} Ohsuga, K. 2007, \apj, 659, 205
\bibitem[Ohsuga et al.(2005)]{O05} 
 Ohsuga, K., Mori, M., Nakamoto, T., \& Mineshige, S. 2005, \apj, 628, 368
 (Paper I)
\bibitem[Okajima et al.(2006)]{Okajima06} 
 Okajima, T., Ebisawa, K., \& Kawaguchi, T.\ 2006, \apjl, 652, L105
\bibitem[Okuda(2002)]{Okuda02}
 Okuda, T. 2002, \pasj, 54, 253
\bibitem[Okuda et al.(1997)]{OFS97}
 Okuda, T., Fujita, M., \& Sakashita, S. 1997, \pasj, 49, 679
\bibitem[Paczy\'nsky \& Wiita(1980)]{PW80} 
 Paczy\'nsky, B. \& Wiita, P. J. 1980, \aap, 88, 23
\bibitem[Parker(1975)]{Parker75} 
 Parker, E.~N.\ 1975, \apj, 198, 205 
\bibitem[Sakimoto \& Coroniti(1989)]{SC89} 
 Sakimoto, P.~J., \& Coroniti, F.~V.\ 1989, \apj, 342, 49
\bibitem[Shakura \& Sunyaev(1973)]{SS73} 
 Shakura, N. I. \& Sunyaev, R. A. 1973, \aap, 24, 337
\bibitem[Shakura \& Sunyaev(1976)]{SS76} 
 Shakura, N.~I., \& Sunyaev, R.~A.\ 1976, \mnras, 175, 613
\bibitem[Shibazaki \& H\=oshi(1975)]{SH75} 
 Shibazaki, N. \& H\=oshi, R. 1975, PThPh, 54, 706
\bibitem[Stella \& Rosner(1984)]{SR84} 
 Stella, L., \& Rosner, R.\ 1984, \apj, 277, 312 
\bibitem[Turner et al.(2005)]{Turner05} 
 Turner, N.~J., Blaes, O.~M., Socrates, A., Begelman, M.~C., 
 \& Davis, S.~W.\ 2005, \apj, 624, 267 
\bibitem[Turner et al.(2003)]{Turner03} 
 Turner, N. J., Stone, J. M., Krolik, J. H., \& Sano, T. 2003, \apj, 593, 992
\bibitem[Vierdayanti et al.(2006)]{Vierdayanti06} 
 Vierdayanti, K., Mineshige, S., Ebisawa, K., \& Kawaguchi, T. 2006,
 \pasj, 58, 915
\bibitem[Wang et al.(1999)]{Wang99} 
 Wang, J.-M., Szuszkiewicz, E., Lu, F.-J., \& Zhou, Y.-Y.\ 1999, \apj, 522, 839
\bibitem[Watarai \& Mineshige(2003)]{WM03} 
 Watarai, K., \& Mineshige, S. 2003, \pasj, 55, 959
\bibitem[Watarai et al.(2001)]{WMM01} 
 Watarai, K., Mizuno, T., \& Mineshige, S. 2001, \apj, 549, L77
\end{thebibliography}
\end{document}